\documentclass[useAMS,usenatbib,psfig]{mn2e}
\usepackage{epsfig}
\usepackage{graphicx}
\usepackage{array}

\title[THE DISC MASS OF SPIRAL GALAXIES]{THE DISK MASS OF SPIRAL GALAXIES}
\author[P.Salucci, I.A.Yegorova,  N. Drory]{Paolo Salucci$^{1}$\thanks{E-mail:
salucci@sissa.it}, Irina A. Yegorova$^{1}$ \& Niv Drory$^{2}$
\\\\
$^{1}$SISSA  International School for Advanced Studies, via Beirut 4,
I--34013 Trieste, Italy\\
$^{2}$Max-Planck Institut f\"ur extraterrestrische Physik, Giessenbachstra\ss e, Garching, Germany}
\begin{document}


\pagerange{\pageref{firstpage}--\pageref{lastpage}} \pubyear{2002}

\maketitle

\label{firstpage}

\begin{abstract}

  We derive the disk masses of 18 spiral galaxies of different
  luminosity and Hubble Type, both by mass modelling their  rotation curves and by
  fitting their SED with spectro-photometric models.  The good agreement of 
  the estimates obtained from these two different methods allows us to quantify 
  the reliability of their performance
    and   to derive very  accurate
  stellar mass-to-light ratio vs color (and stellar mass) relationships.
  
\end{abstract}


\section{Introduction}

The disk mass $M_D$, together with disk length-scale $R_D$, is the
main physical property of the baryonic component of normal spiral
galaxies. In the current framework of galaxy formation theory, in an
halo of mass $M_{vir}$, the present day value of $ M_D$ indicates the
efficiency with which the stellar formation process acted in
proto-spirals on the large primordial reservoir of $\sim {1\over{6}}
M_{vir} $ HI material. It also  bears the imprint of the physical processes
(supernovae feedback, cooling, previrialization) that have prevented
the latter in entirely transforming into a stellar disk (see Shankar et
al. 2006). The quantity $(M_D R_D)^{1/2}$ is proportional to the angular 
momentum  per unit mass of the  disk matter, very likely the same value of   
that of the dark particles (e.g. Tonini et al, 2006) and it is linked to the tidal
torques that dark halo  experienced  from neighbors at their turnarounds.  
Finally, we should recall that $ M_D R_D^{-2} $ is a measure of the central 
stellar surface density.
   
The mass modelling of the rotation curves (hereafter RCs) is a robust
and reliable method (hereafter   the kinematical method 
{\it kin})   to derive the disk mass (e.g. Tonini \& Salucci, 2004)). 
For  illustrative purposes  we recall  that, inside one disk length-scale  
the  RC's almost entirely  match the distribution of the stellar component,  a perfect match between gravitating mass and luminous mass  is reached by adding just a very small  dark matter component (e.g. Persic, Salucci, Stel  1996, Salucci \& Persic, 1999).  This  observational evidence  
is well understood    within the current  galaxy formation scenario. Disks form from  baryons  that,  while infalling in   DM
potential wells,     unlike the DM component, radiate, lose kinetic energy and
contract: their  final configuration  saturates the gravitational field of the inner
regions of galaxies.

A second independent way, pioneered by Tinsley (1981)  (hereafter  the
spectro-photometric method  {\it pho}),    by fitting  a galaxy 
broadband SED  with stellar population models obtains the   disk mass as the resulting   best-fit parameter.

In  Salucci et al. 1991 we find  the  first study  in which the   combination of these two independent   methods      
was applied to a sample of 38 spirals   with available kinematics and photometry;   disk masses were estimated  from the $B-V$ colors 
and  then compared with the values obtained by modelling their  rotation curves.   
Similarly,  Ratnam and Salucci, (2000)  used high-spatial resolution ($<$ 100 pc) 
rotation curves of 30 spirals   in order to derive the mass distribution in  their innermost kpc. 
They found that, in this region  the luminous matter completely accounts for the gravitational potential,  so that  the kinematics provides a very precise value for the  stellar mass-to-light ratio.  These values   resulted   in  a good agreement   with that obtained from $B-I$ colors by applying   the predictions of  population synthesis models. A similar result has been found for  a small number of  "laboratory" barred spirals (Weiner et al , 2001; Perez, 2004)

A substantial improvement of the {\it pho }  method  has come from  Bell and de Jong, 2001;  they devised spiral galaxy evolution models that yielded     the  stellar M/L ratios   dependence  on  different  colors  for integrated stellar populations.
These results were tested against the  "maximum disk"  stellar M/L ratios of a sample of spiral galaxies.

In a different approach  the  determination  of disk masses was directly implemented in the 
RC mass modelling itself.   Kassin et al. (2006) studied a sample of 34 bright spiral galaxies whose disk  masses 
were obtained by   {\it BRK } colors,  so that  the mass model had one less  free parameter. 
However,  this promising procedure requires  the mass estimate to  be  very accurate;   
since   an  error on  the assumed value $M_{pho}$, with respect to the actual value $M_{true}$,  
fatally flaws the  entire mass modelling (e.g. Tonini and Salucci, 2006). It is useful  now to  give  a   simplified proof of this. 
Let us set   $V_h \propto R^s $, the halo velocity contribution to the circular velocity around $2.2 R_D $, the radius where the disk contribution to the circular velocity (of true value $\beta\simeq V^2_d/V^2$)  has a  radial maximum.   By setting   $\nabla$ the (observed)  RC slope  at $2.2 R_D $,   we have: $ s= \nabla/(1-\beta \ M_{pho}/ M_{true})$ that shows   even    quite small  errors   in the {\it pho } estimate of  $M_{true}$ may trigger    very  large errors in   derived DM  halo profiles.

Advantages and disadvantages are present in both {\it pho} and {\it  kin}  methods, though,
remarkably, they are almost orthogonal. The photometric method relies
on the not trivial caveat that, given a SED there is an unique stellar
mass which explains it. Moreover, it takes a number of assumptions on
the stellar populations of spirals, it depends on the estimate of the
galaxy distance as $D^2$ (uncertain to some level for local
objects). Moreover, it carries theoretical uncertainties    (in log
$M_D$) up to 0.3 dex. The kinematical method may suffer from
the uncertainty on the actual distribution of DM in galaxies, depends
on the disk inclination angle as $(1/ \sin i)^2$ and   therefore   is  uncertain for low
inclination galaxies, and it depends on the estimate of the galaxy
distance as $D$. The main advantages of the photometric method  are  that
it  makes  no assumption on the mass distribution and it requires only a limited observational effort; 
the main advantage of the
kinematical method is that the disk mass can be obtained
straightforwardly from  model independent observables and within an uncertainty of $0.15 dex$: 
infact $M_D \sim f G^{-1} V^2(2R_D) 2 R_D$, where $f$ can be estimated from 
the slope of the RC inside $2 R_D$, within 0.1 dex  uncertainty 
(Persic and Salucci, 1990a,b),  while  $R_D$ is  generally known within a 0.07 dex  uncertainty.

Combination of the two methods  is particularly useful in dark matter dominated dwarfs where the {\it kin } method fails since the stellar disk is largely non self-gravitating  at any radius and 
in galaxies with  large uncertainties in  the distances  that propagate themselves  onto the  {\it pho } estimates.

The aim of this paper is to measure the  disk masses of a sample of 18  spirals with both 
methods and to compare them. We use combined SDSS and 2MASS photometry covering the \textit{ugrizJHK}
bands to fit each galaxy's SED to a broad range of stellar population
models with  varying star formation history, age, dust content, and burst
fraction. With respect to previous {\it pho} estimates  the present one takes advantage 
of measurements in 8 bands ranging from the  u to  the  K. 
 High quality RC's will provide  the kinematical data, 
 whose analysis will benefit from  the results of recent work.   

Comparing, for a fair sample of galaxies, the disk mass estimates
obtained with the two different methods is worthwile for the following reasons: 
firstly, the average of the two estimates will give a very reliable
measure of the stellar disk mass for objects of different luminosities, 
thus providing an unprecedentedly accurate stellar mass vs light relationship.  
Secondly, it allows for an  investigation of the assumptions taken by each of the two 
methods,  providing   additional information on the structure and evolution 
of galaxies. Finally, it will indicate   the still poorly known  uncertainties
of  the   {\it pho}  estimates  of the disk mass, setting the  situation in which they can be used in 
  mass modelling.   

The sample consists of 18 galaxies (see Fig 1 for their images). Among them 16 galaxies are late-type
spirals taken from Courteau (1997) and Vogt (2004). The data analysis
technique applied to these rotation curves  has been described in Yegorova et al.
(2007);  in order  to  also include    some  early type objects  we decided to add  UGC4119 and   UGC6351 whose  data were obtained
in Asiago Observatory with the 182cm Copernico telescope in January 2006.
The sample includes all the SDSS local spirals with 8 band  
measurements (i.e. a good coverage of  the  entire SED) necessary 
for the highest  precision estimate of the "photometric"  mass and with   smooth, symmetric,  regular,  high resolution ( $>$ 10 independent data inside two disk length-scales) rotation curves.   Some of the latter  requirements  are implemented  to make us  sure that  the  selected RC's are not affected  by non-axisymmetric features, such as  bars and  spiral structure that, in any case,  affect the RC  slopes rather than the RC amplitudes (that  plays the  major role in the  {\it kin} estimates). Finally, the values of $R_D$ 
 are taken   from  Courteau (1997) and   Vogt (2004),  including  their  inclination angles $i $,  that being relatively   highly  inclined $i \sim  60^0$,    do not affect  their {\it kin}  estimates.    
    
Notice that  in the literature {\it kin}  disk masses have been obtained  for many other
 objects that,  unfortunately,  do not have the set of  photometric data sufficient for 
the specific aim of this work.

\begin{figure}   
\centerline{
\psfig{file=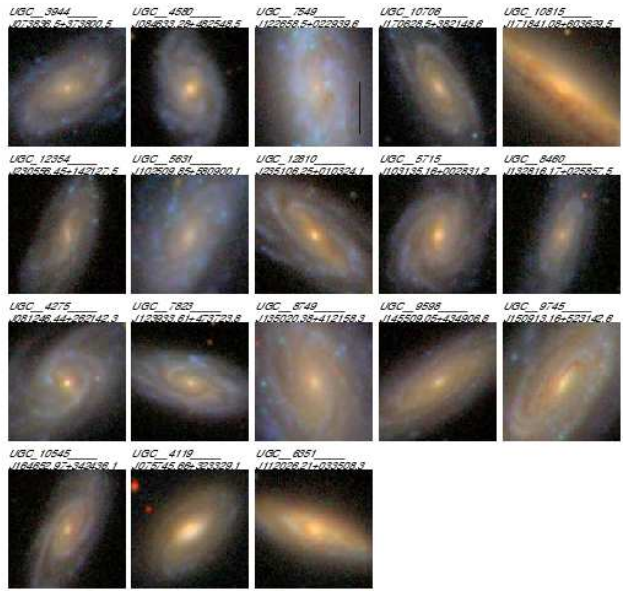,width=8.4truecm}
 }  
 
\vskip 0.1cm  

\caption{SDSS gri composite color images of the galaxies in our sample in the same order of Table 1}
 
\end{figure}   

The plan of this work is the following: in section 2 we will
describe the kinematical method, in section 3 the spectro-photometric
method, the resulting mass and mass-to-light ratio will be shown in
section 4 and 5, and a discussion is presented in  section 6.

\section{The kinematical method}

In spirals the stellar component is represented by a Freeman disk
(Freeman 1970) of surface density:
$$
\Sigma_D(r) = {M_D \over 2 \pi R_D^2}\,
e^{-r/R_D}
\eqno(1)
$$
that contributes to the circular velocity $V$ as:
$$
V_d^2(x) = {1\over 2} {GM_D\over {R_D}}  x^2 (I_0K_0-I_1K_1)
\eqno(2a)
$$
where $x=r/R_D$ and $I_n$ and $K_n$ are the modified Bessel functions
computed at $x/2 $. A bulge of mass $( M_b = \epsilon \ M_D$, $\epsilon =
(1/20-1/5)$ concentrated inside $R_b< 1/3 \  R_D$ is often present. The
amplitude and the profile of the RCs for $R>R_b$ is influenced by the
central bulge in a negligible way for $\epsilon< 1/5$. Furthermore, in
the RC mass modelling, even if we  neglect a quite significant stellar
bulge component ($\epsilon=0.2$), we obtain a disk mass value higher
than the actual one,  but matching that of the total stellar mass
$(M_D+M_b)$ (Persic, Salucci, Ashman, 1993), therefore, providing a suitable  mass  
to be compared with the total galaxy luminosity and with the
spectro-photometric mass estimate.

\begin{figure}
\hspace{-0.5cm}
\begin{tabular}{>{\centering}p{2.4cm}>{\centering}p{2.4cm}>{\centering}p{2.4cm}}
\includegraphics[bb=62bp 189bp 496bp 485bp,clip,width=0.159\textwidth]{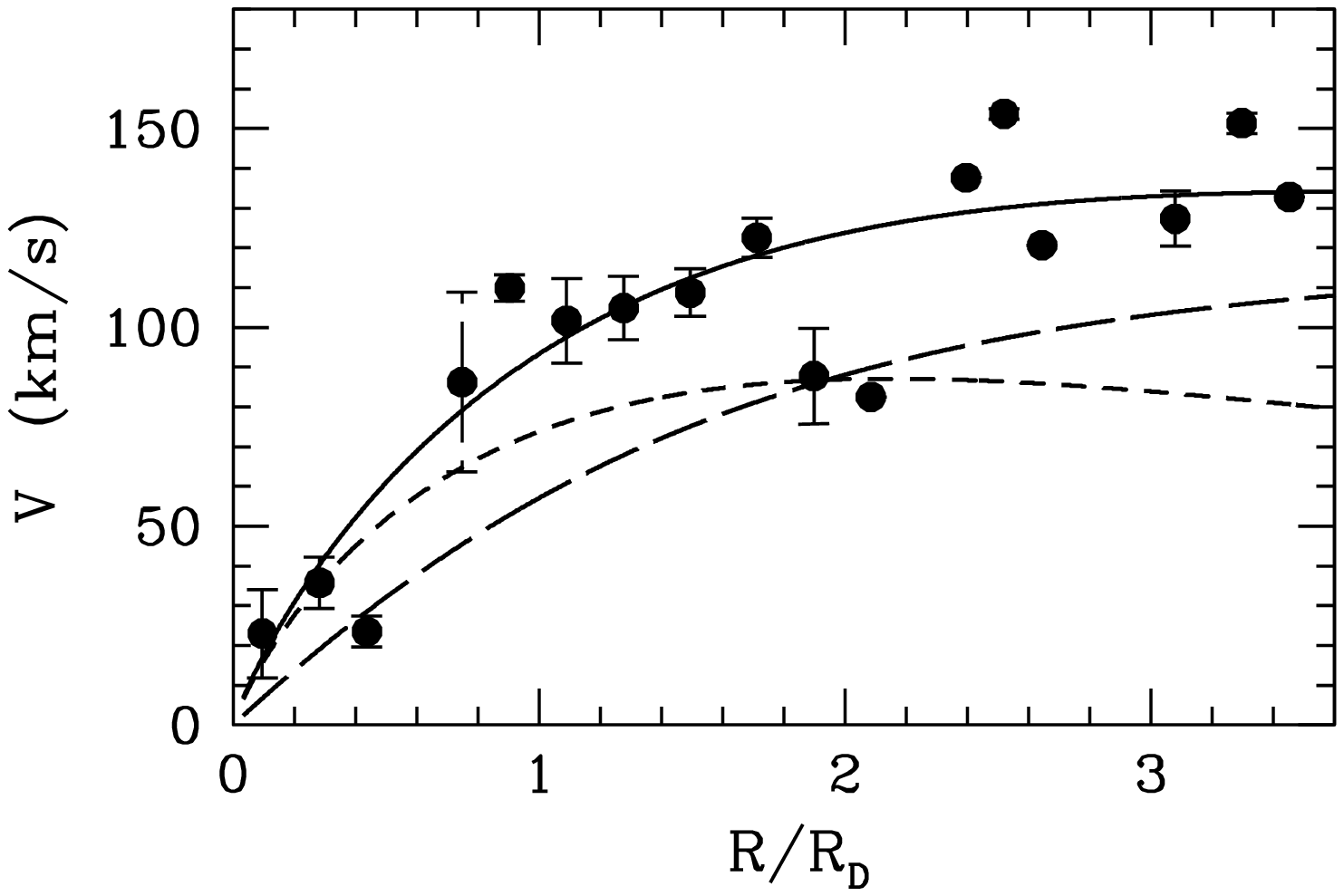}

UGC 3944 & \includegraphics[width=0.16\textwidth]{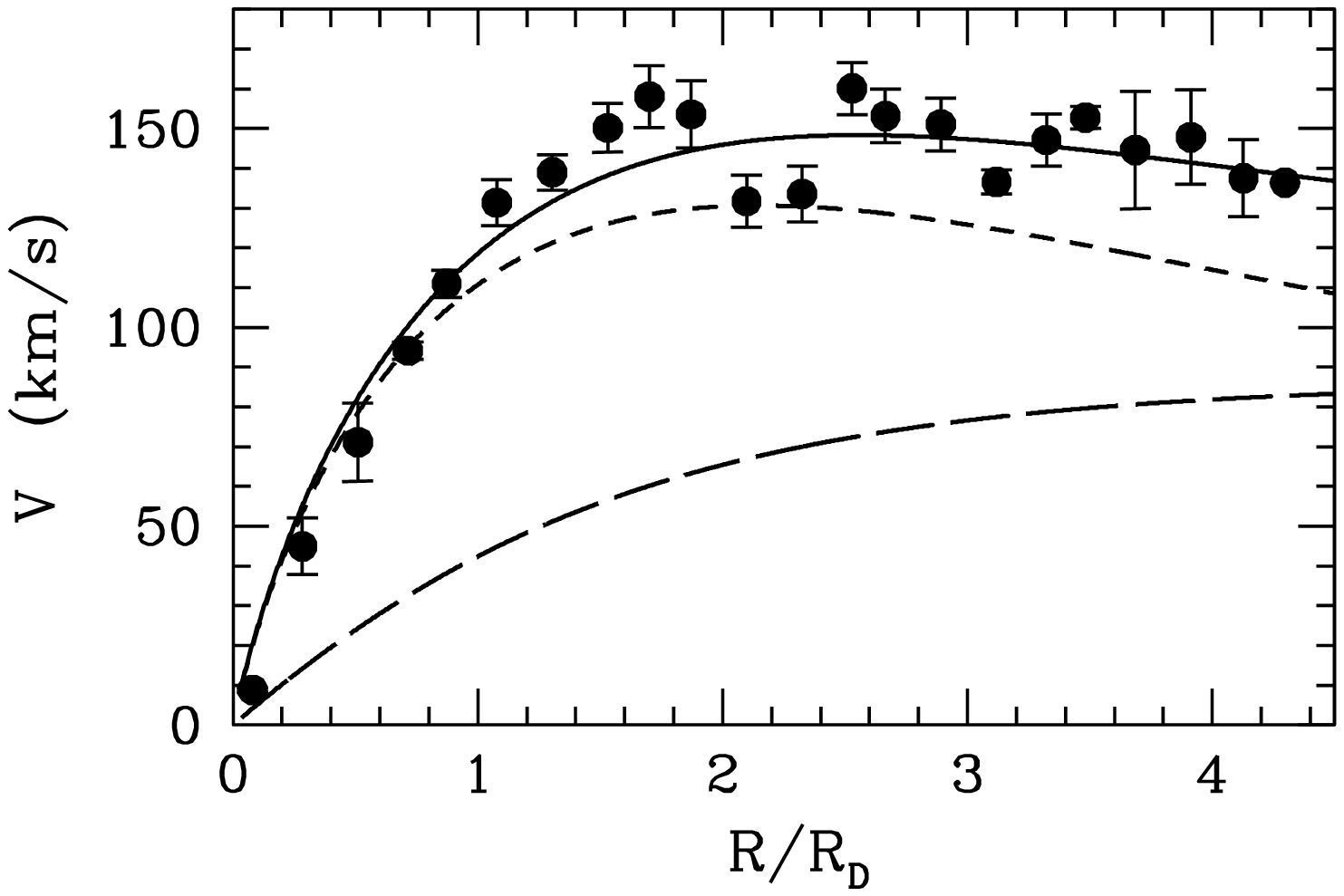}

UGC 4580 & \includegraphics[bb=62bp 189bp 498bp 485bp,clip,width=0.16\textwidth]{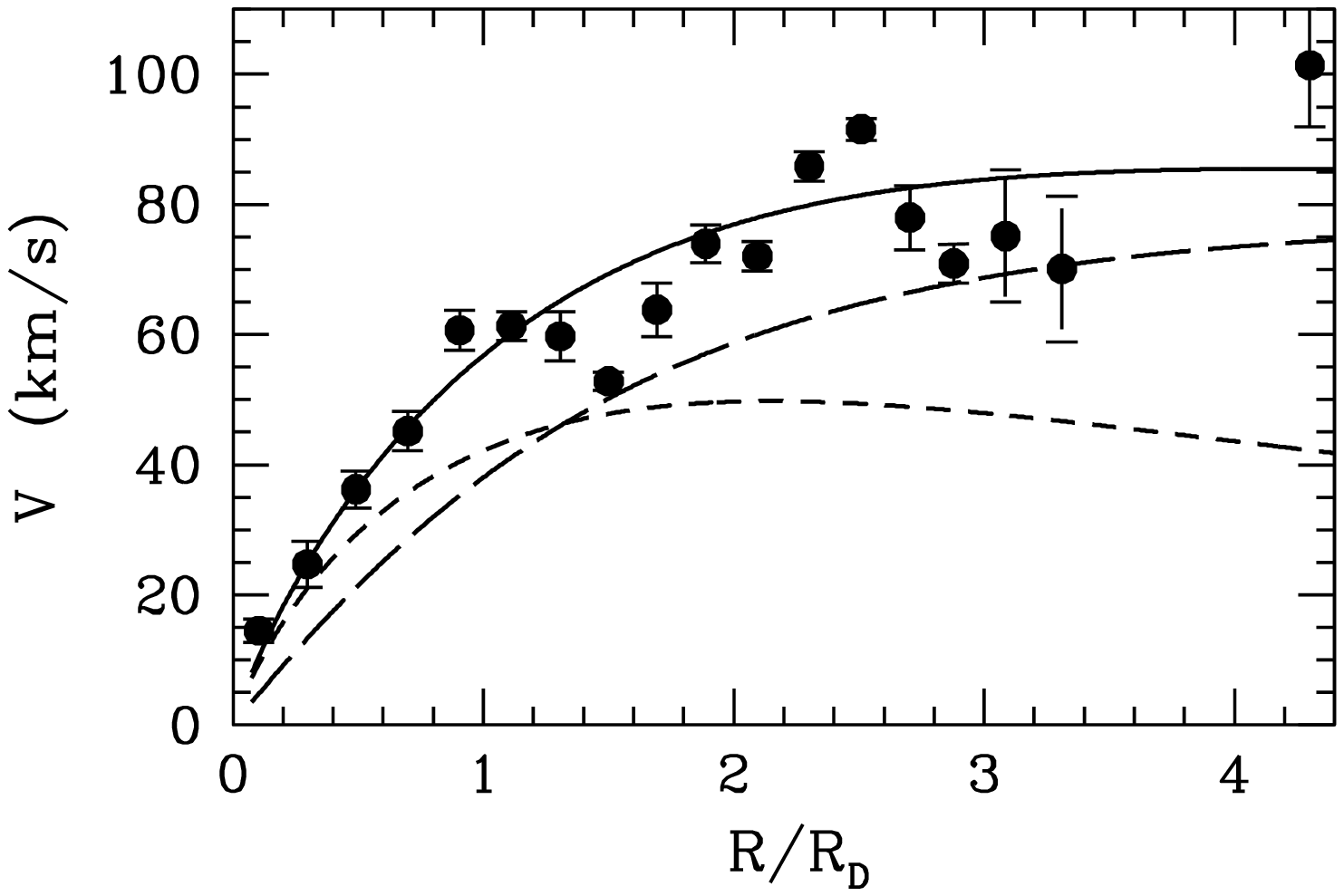}

UGC 7549\tabularnewline
\includegraphics[bb=62bp 189bp 496bp 485bp,width=0.16\textwidth]{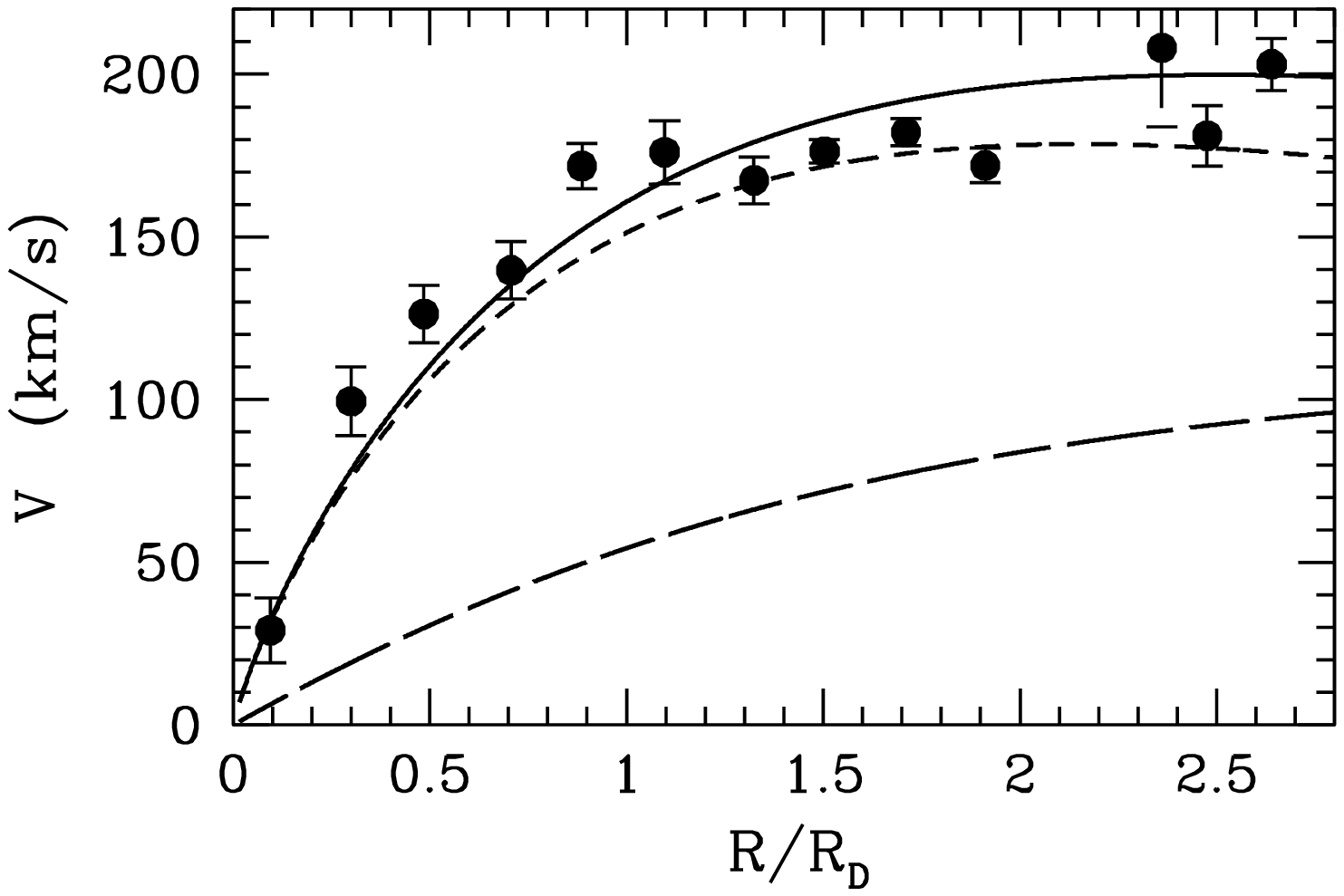}

UGC 10706 & \includegraphics[width=0.16\textwidth]{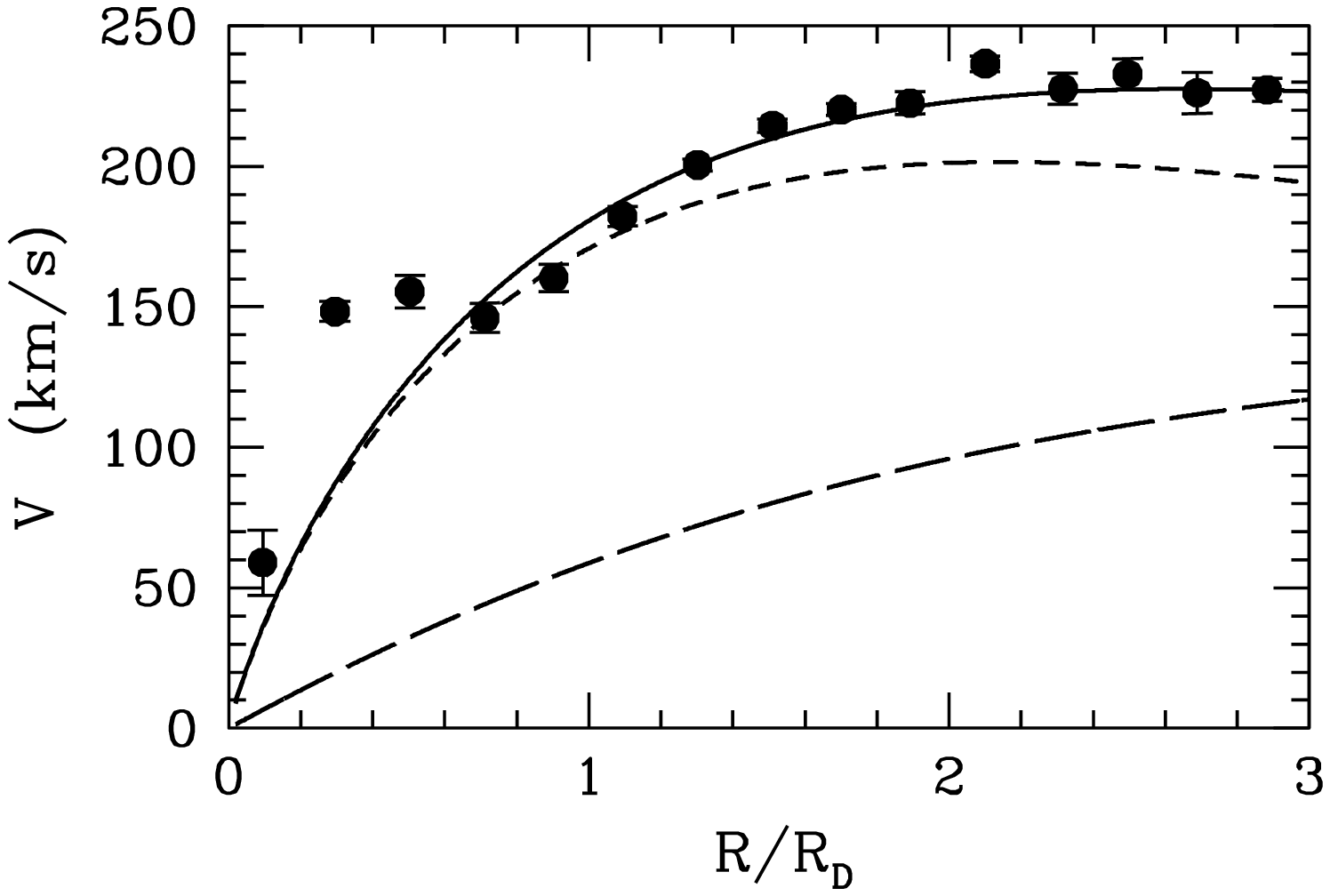}

UGC 10815 & \includegraphics[width=0.16\textwidth]{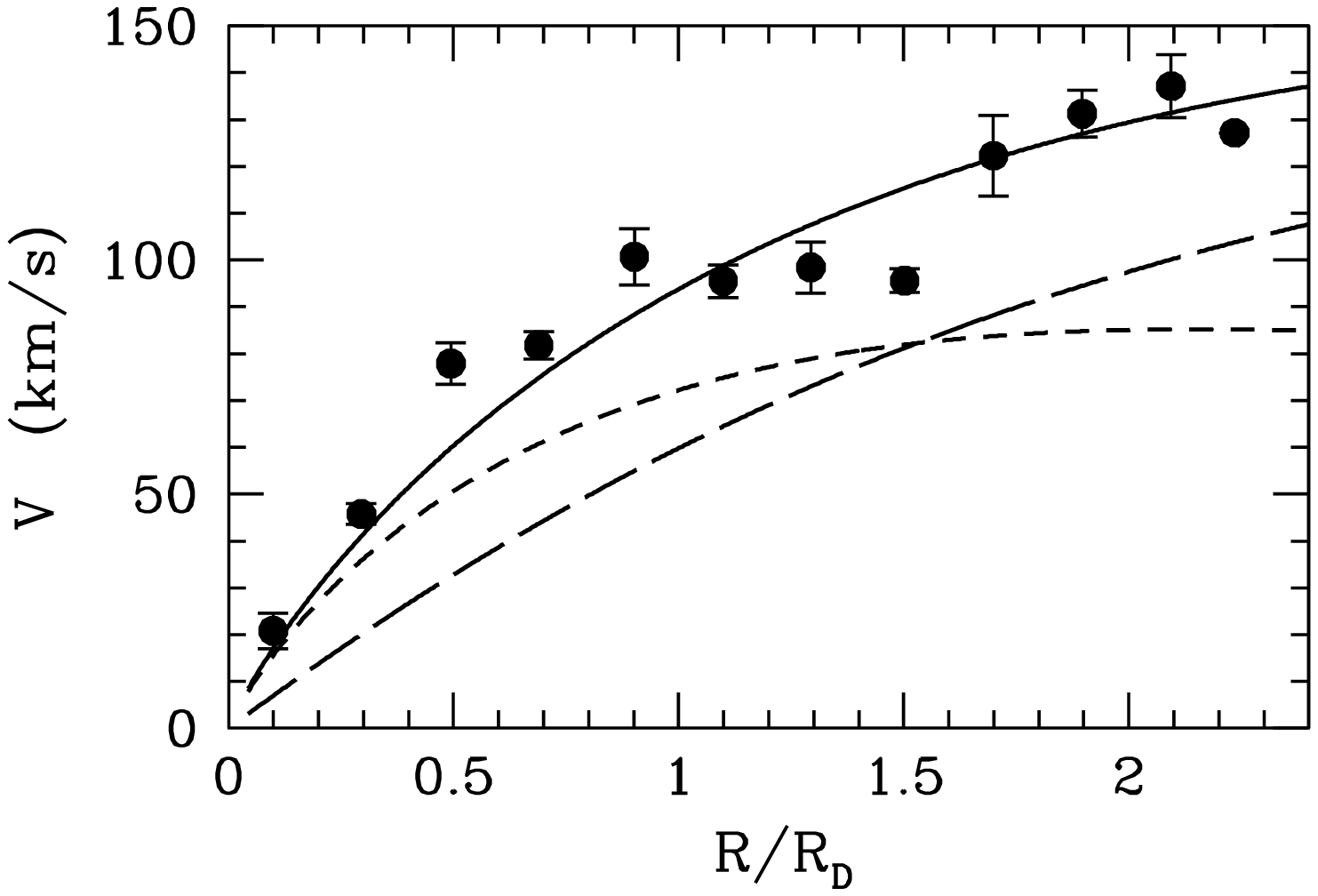}

UGC 12354\tabularnewline
\includegraphics[width=0.16\textwidth]{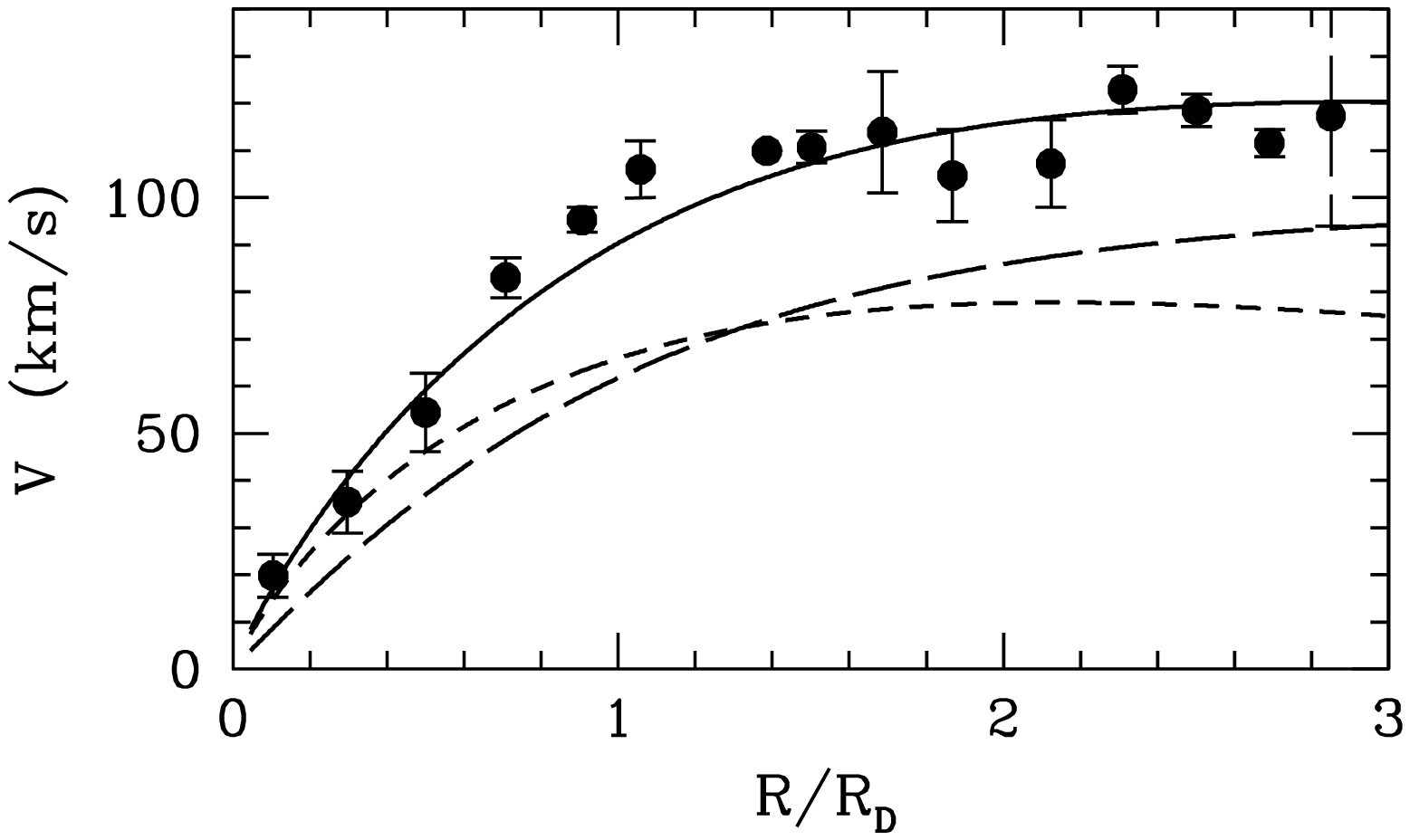}

UGC 5631 & \includegraphics[width=0.16\textwidth]{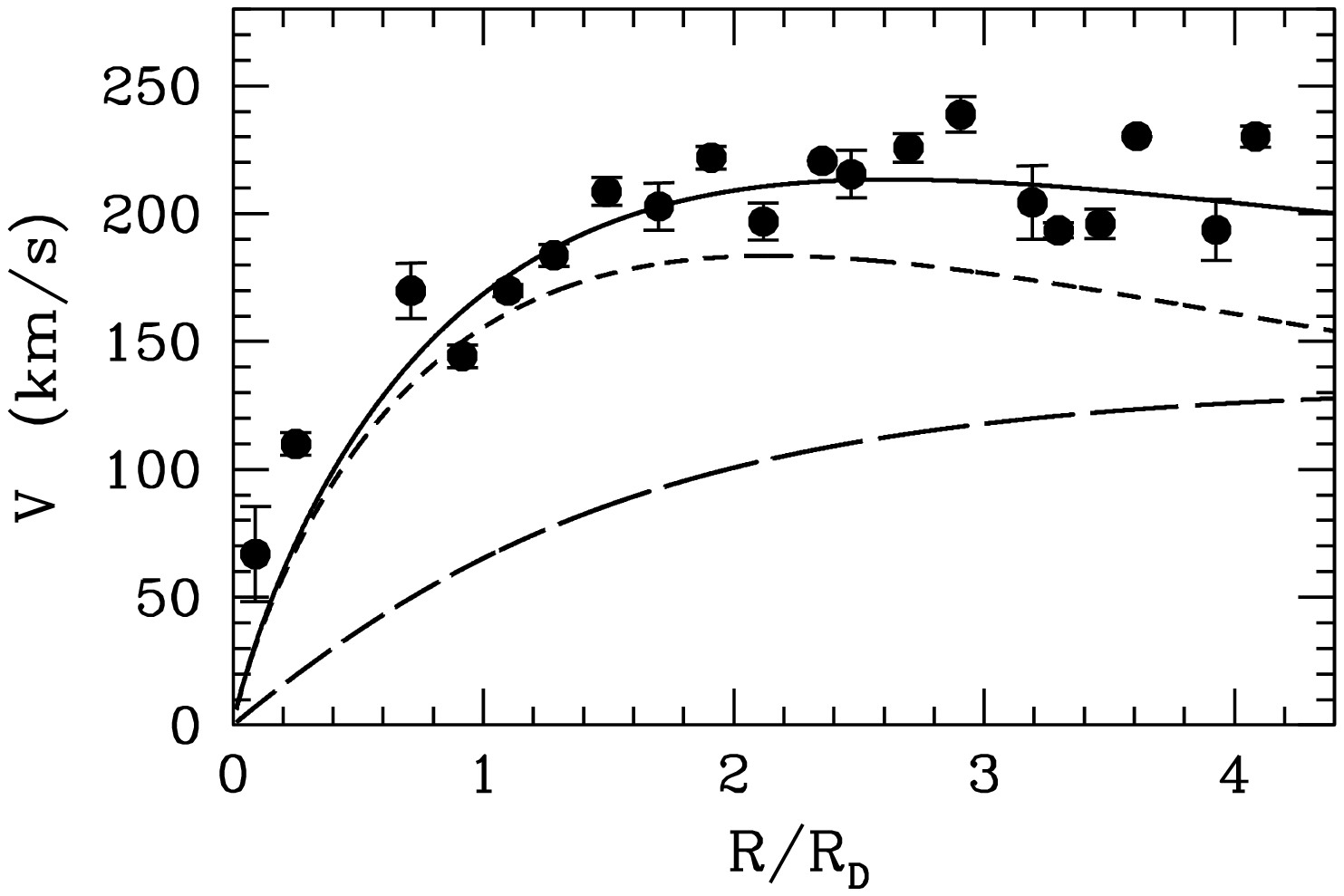}

UGC 12810 & \includegraphics[width=0.16\textwidth]{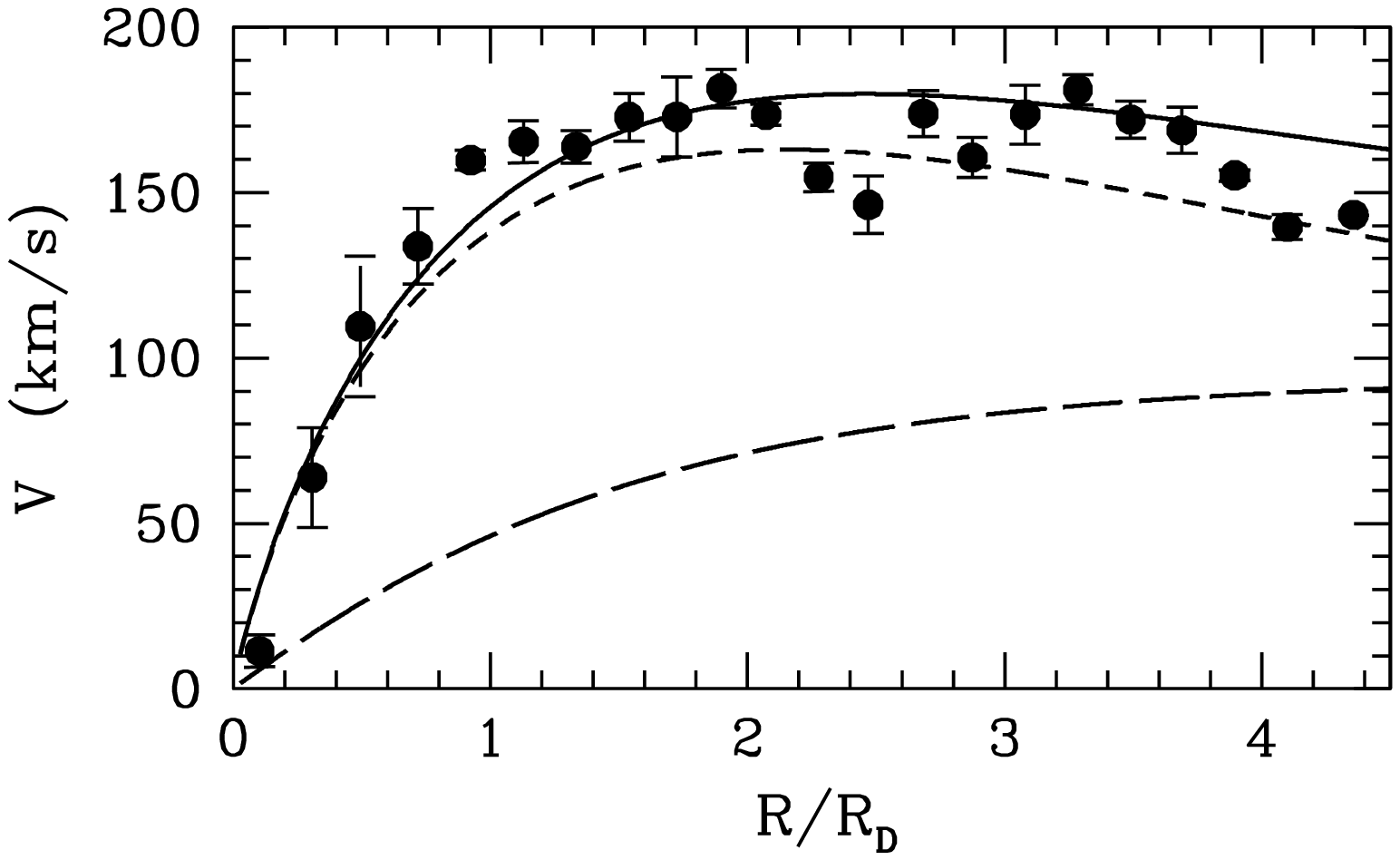}

UGC 5715\tabularnewline
\includegraphics[width=0.156\textwidth]{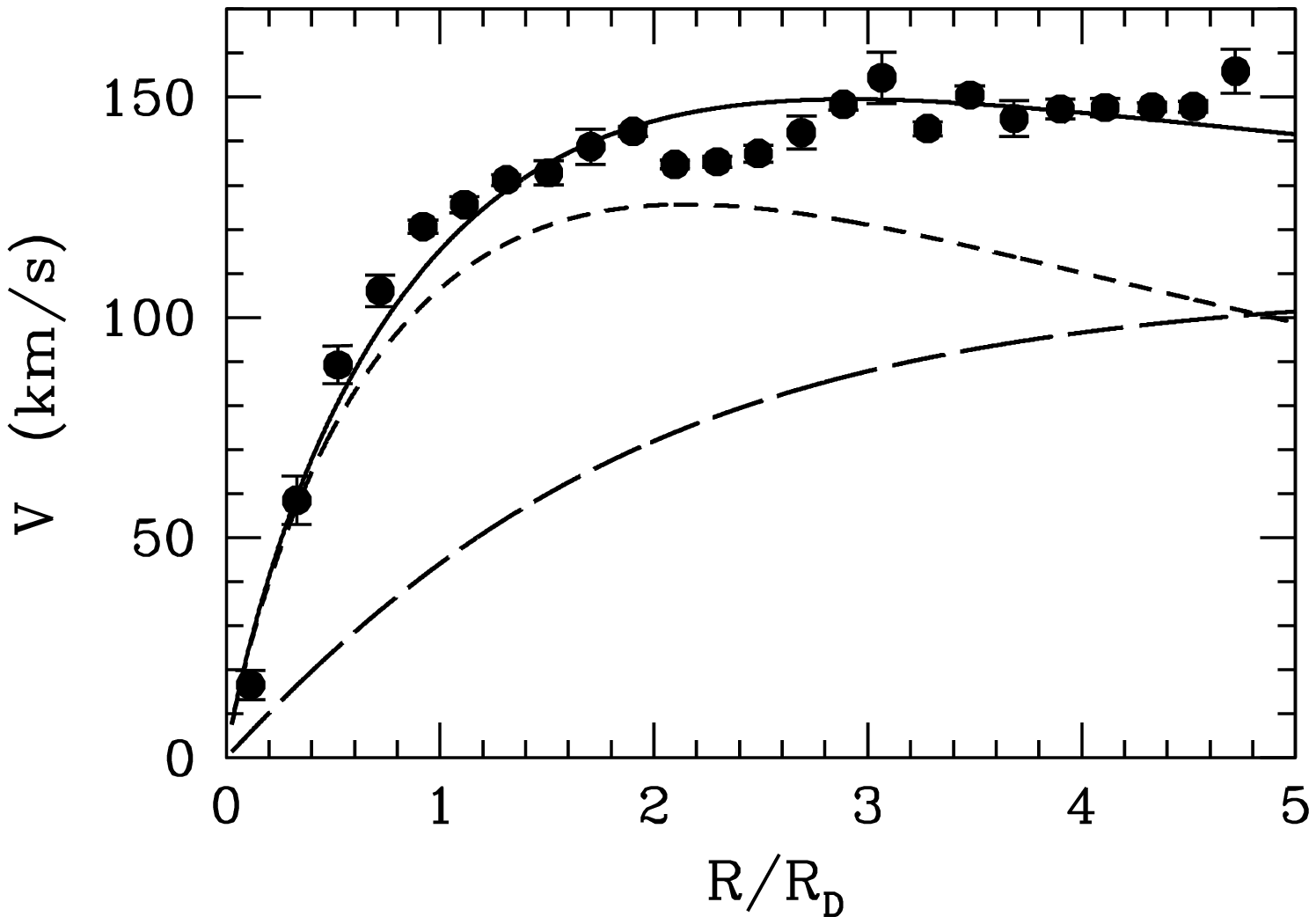}

UGC 8460 & \includegraphics[width=0.16\textwidth]{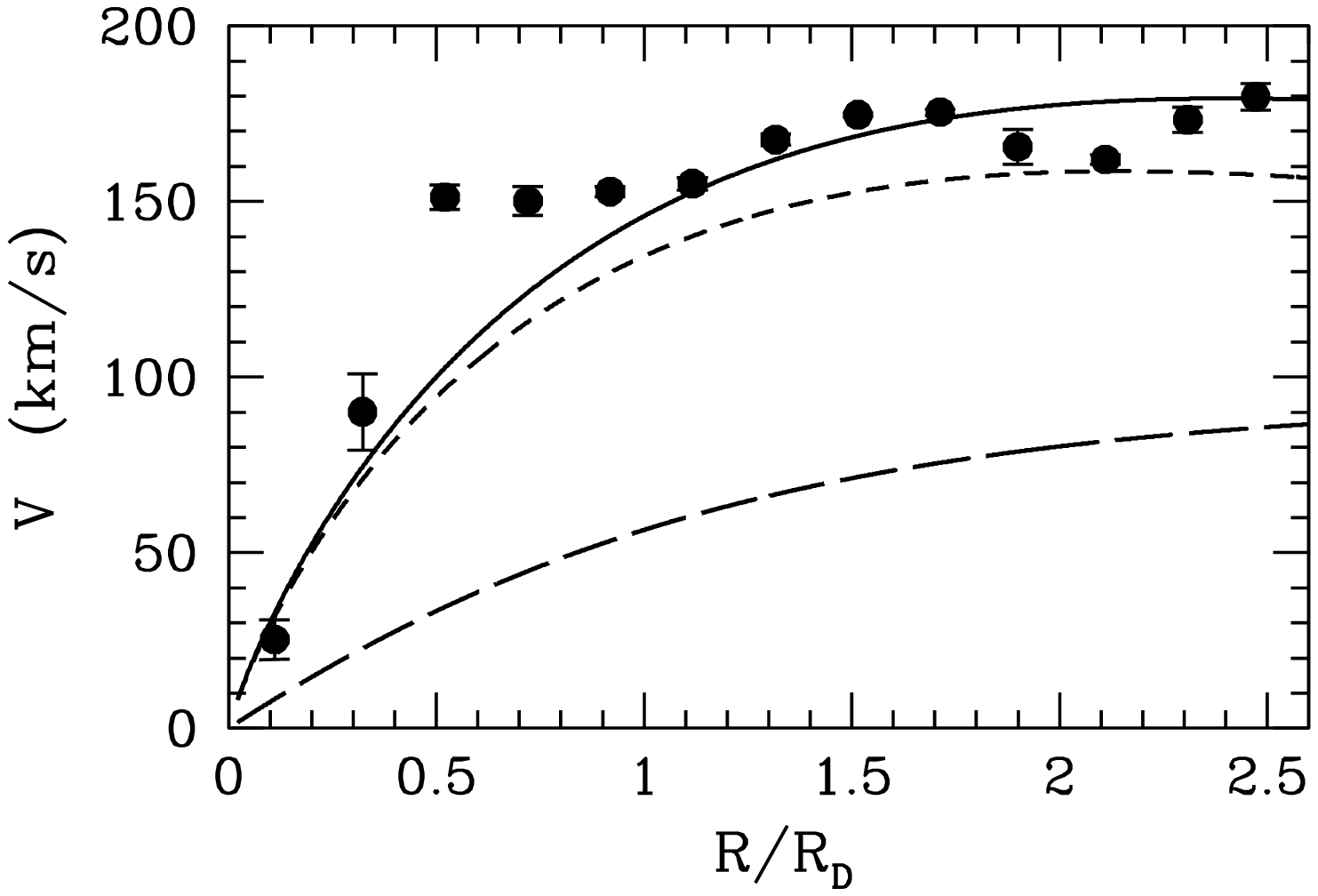}

UGC 4275 & \includegraphics[width=0.16\textwidth]{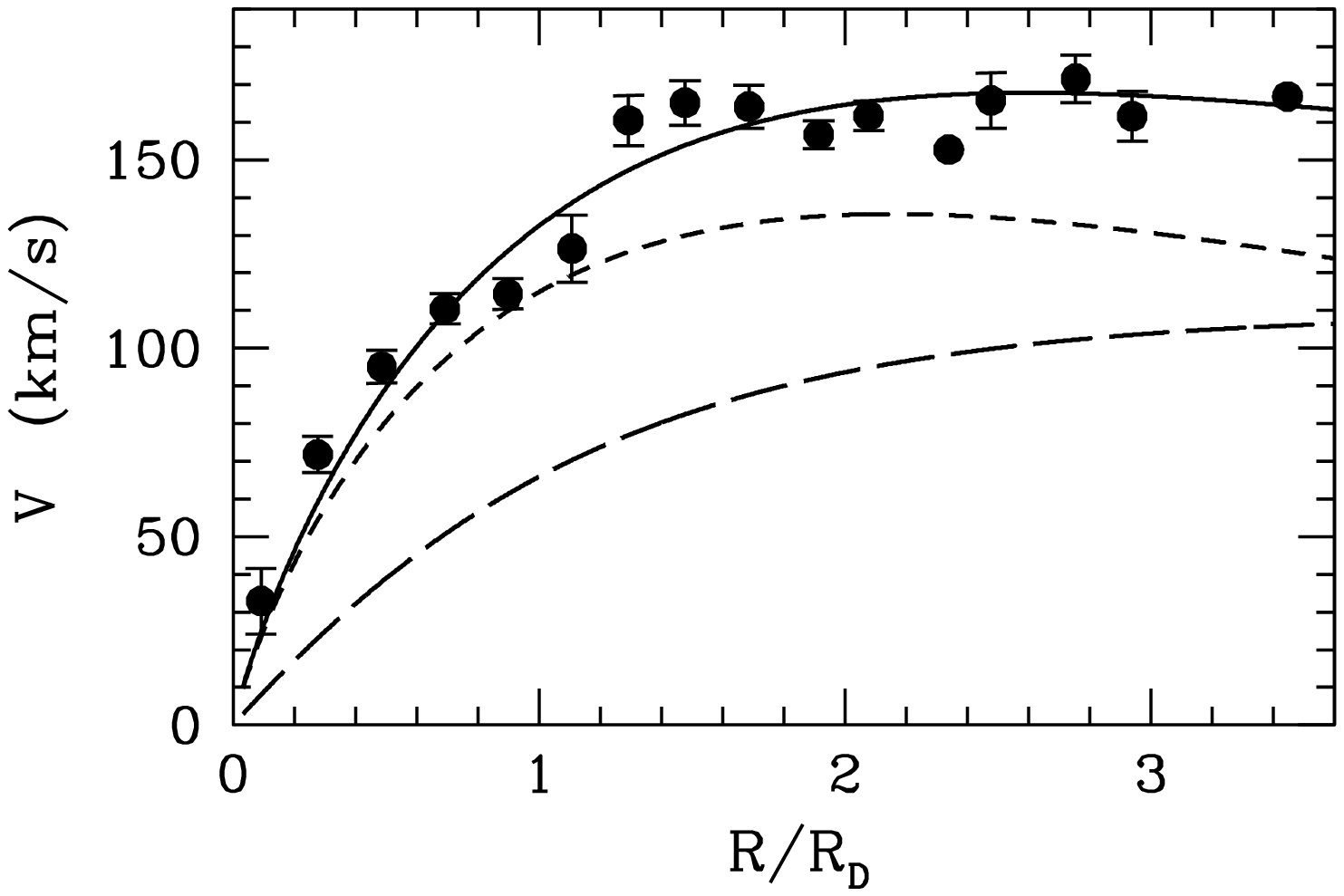}

UGC 7823\tabularnewline
\includegraphics[width=0.16\textwidth]{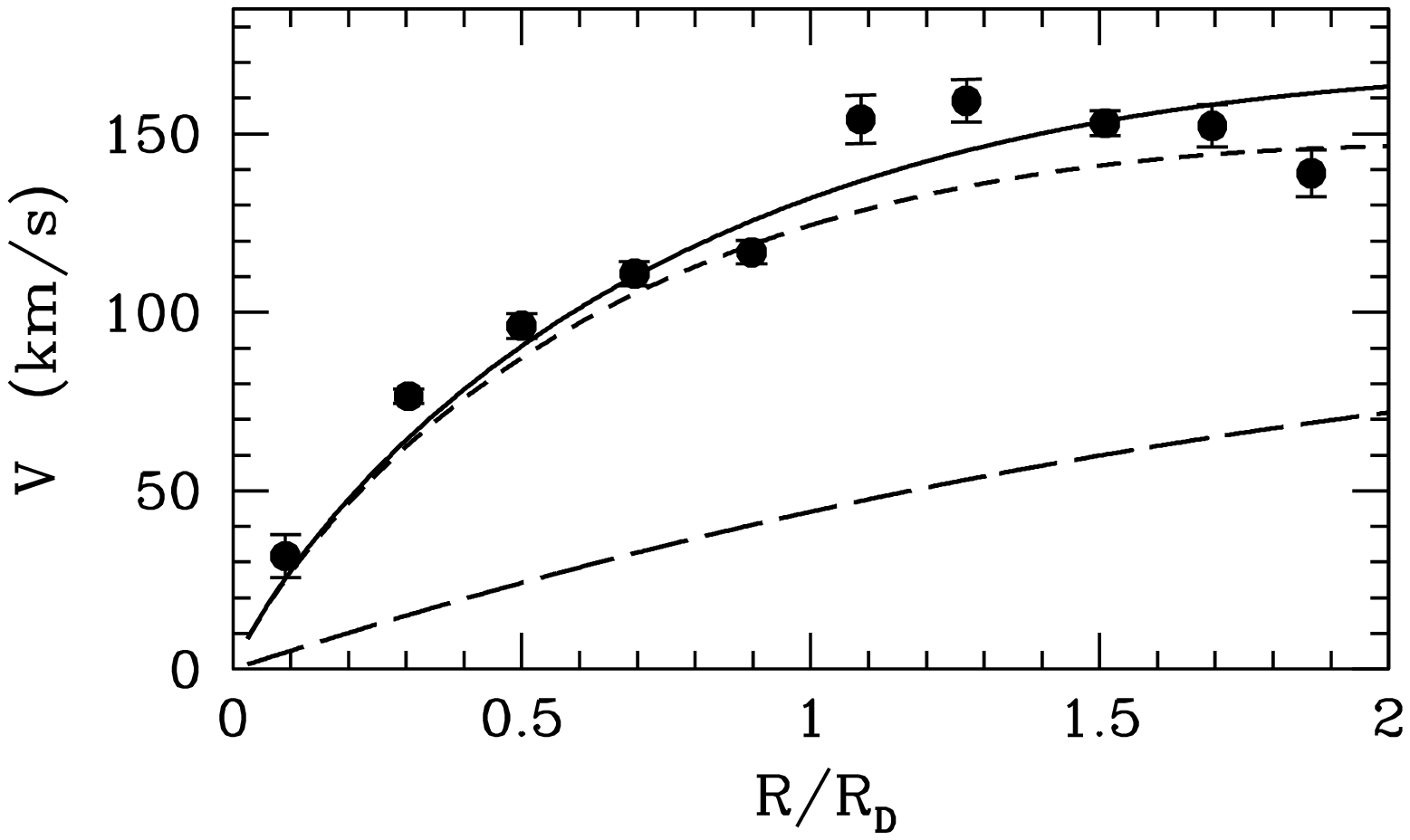}

UGC 8749 & \includegraphics[width=0.16\textwidth]{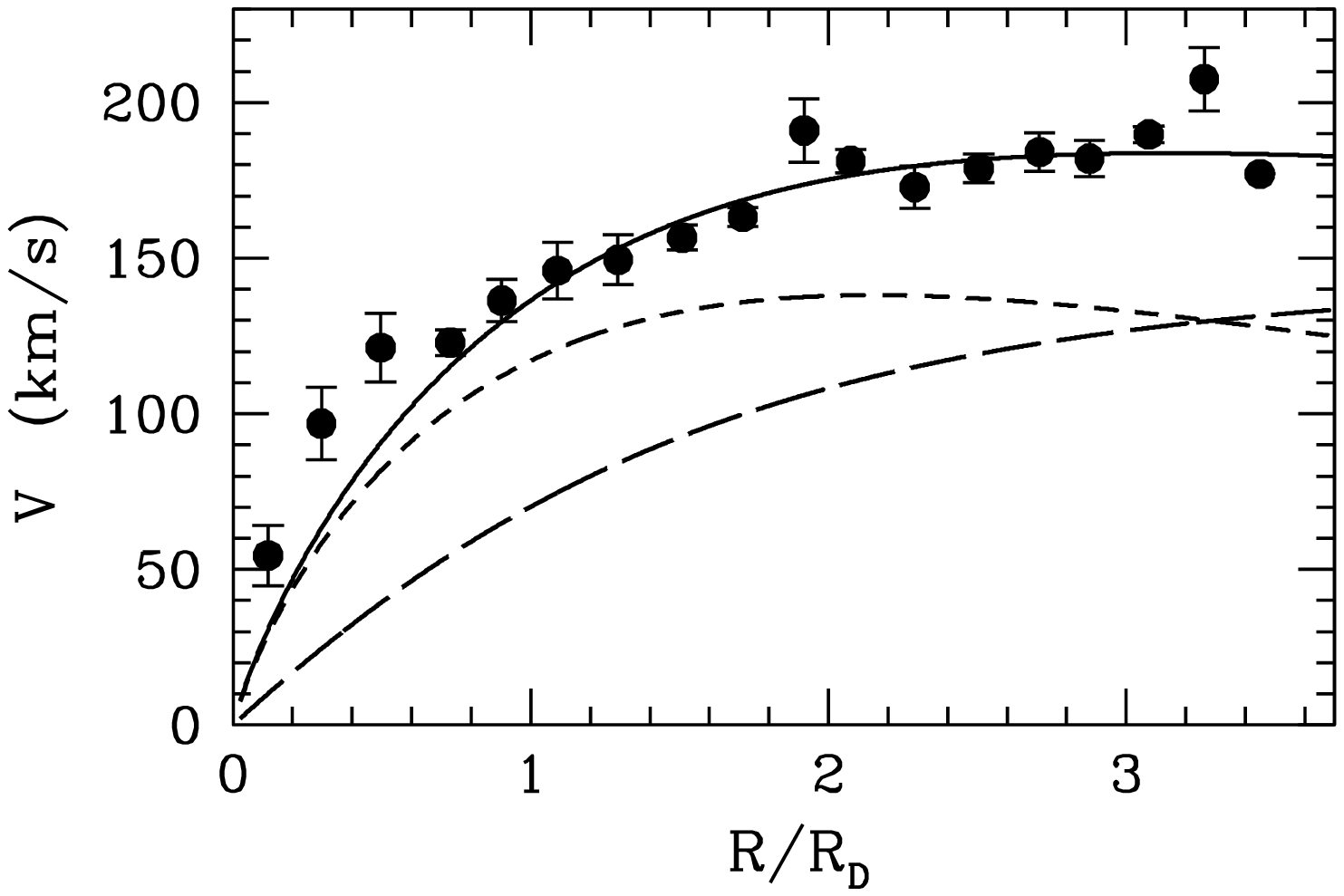}

UGC 9598 & \includegraphics[width=0.16\textwidth]{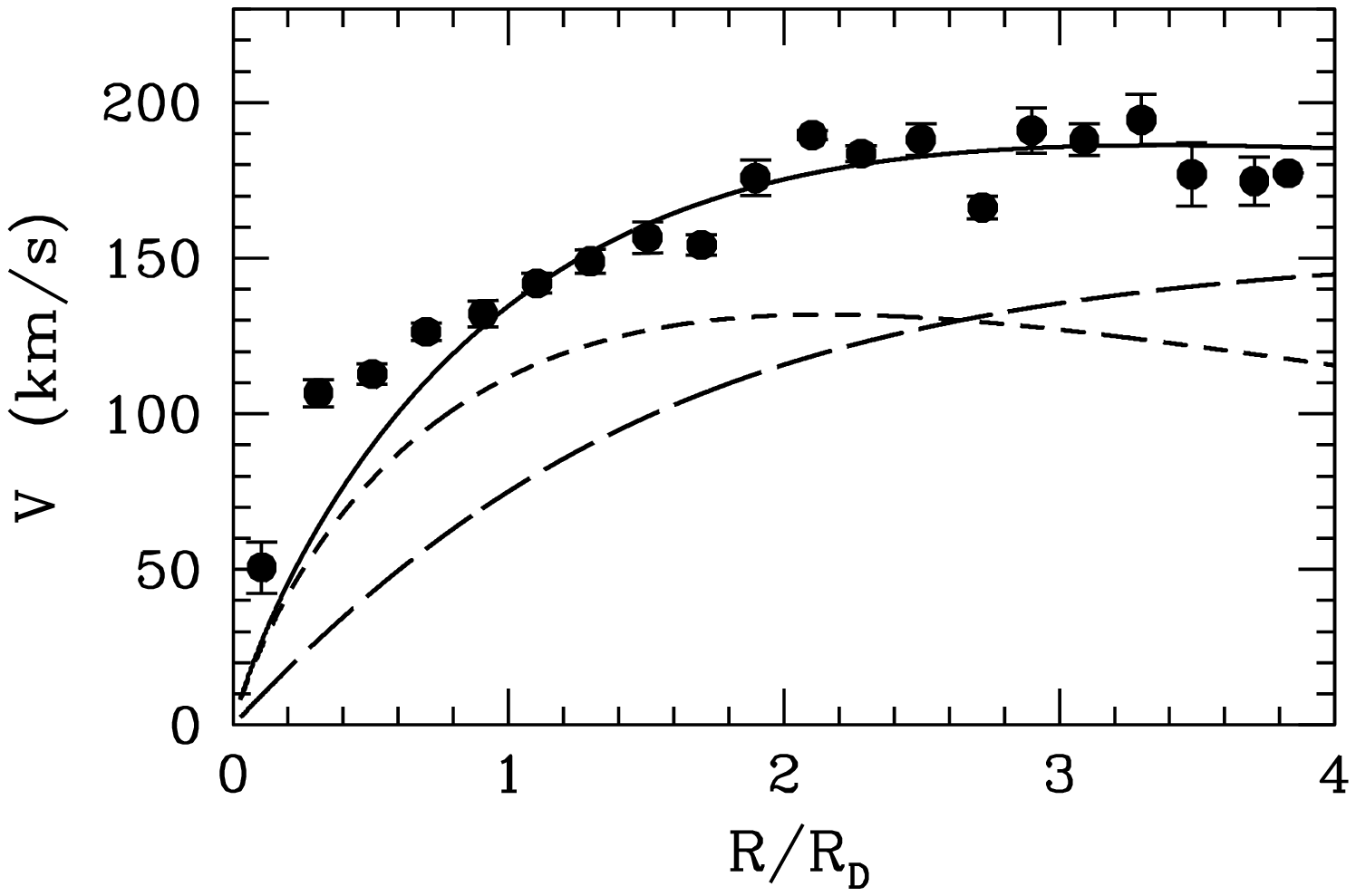}

UGC 9745\tabularnewline
\includegraphics[width=0.16\textwidth]{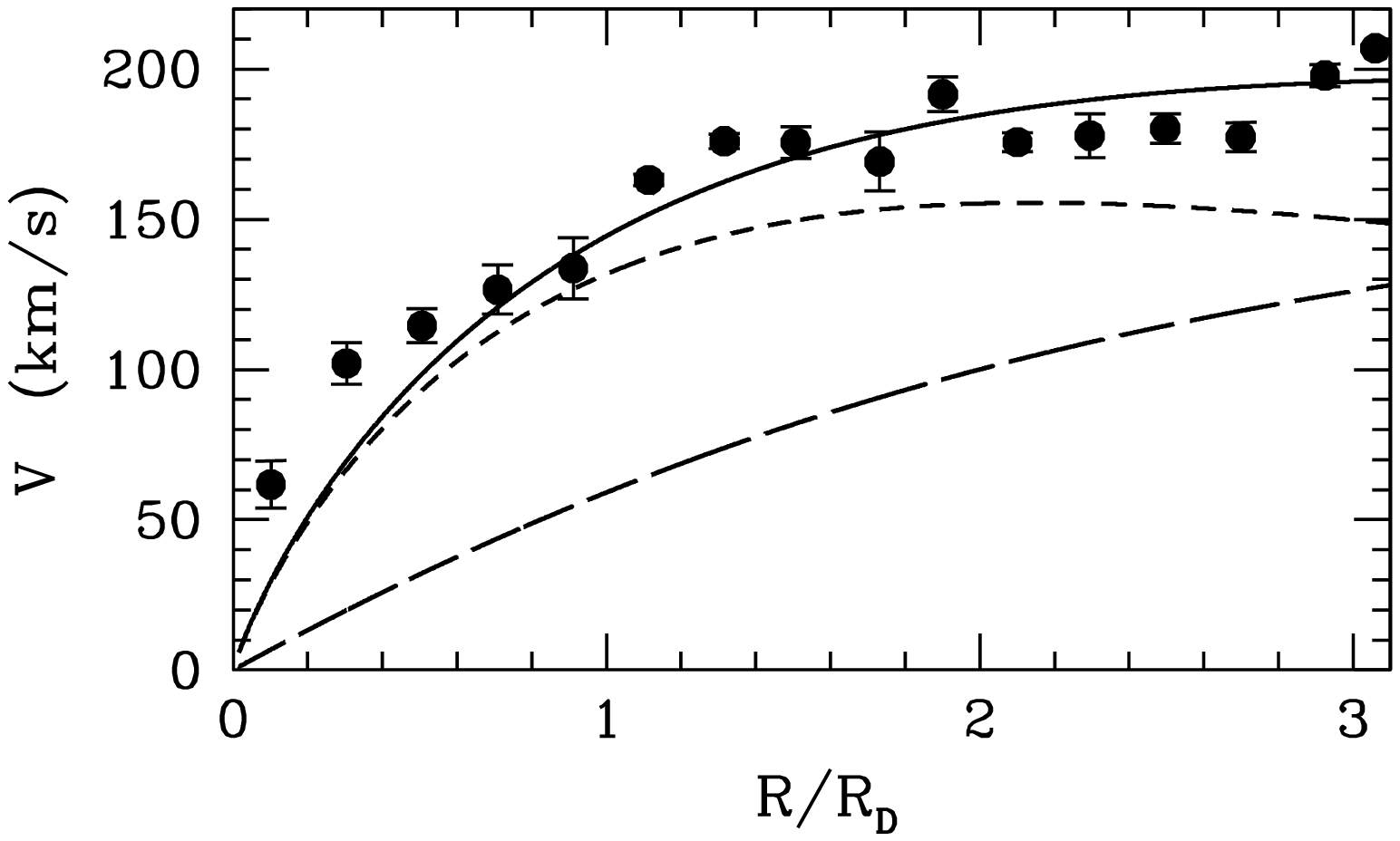}

UGC 10545 & \includegraphics[width=0.16\textwidth]{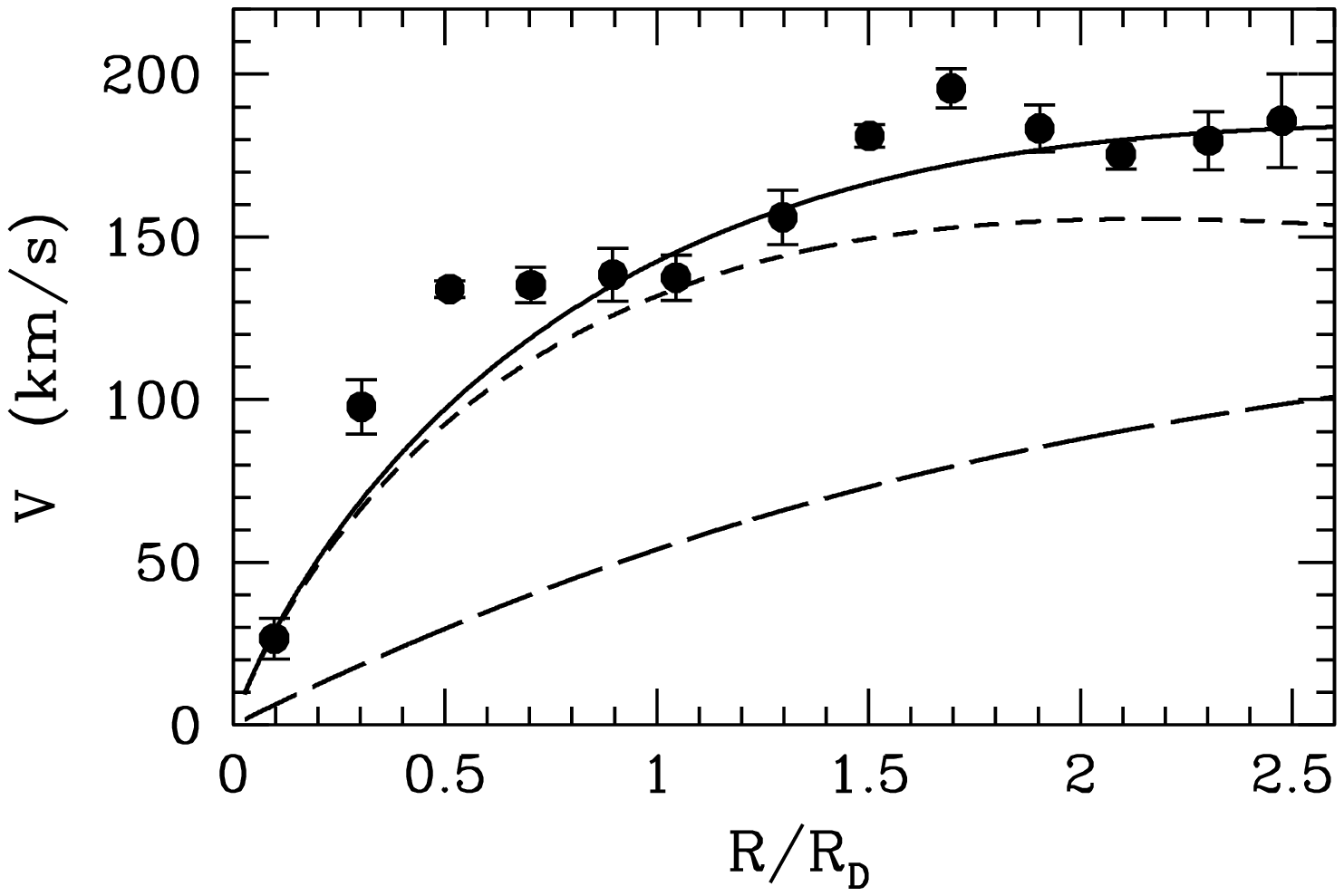}

UGC 4119 & \includegraphics[width=0.16\textwidth]{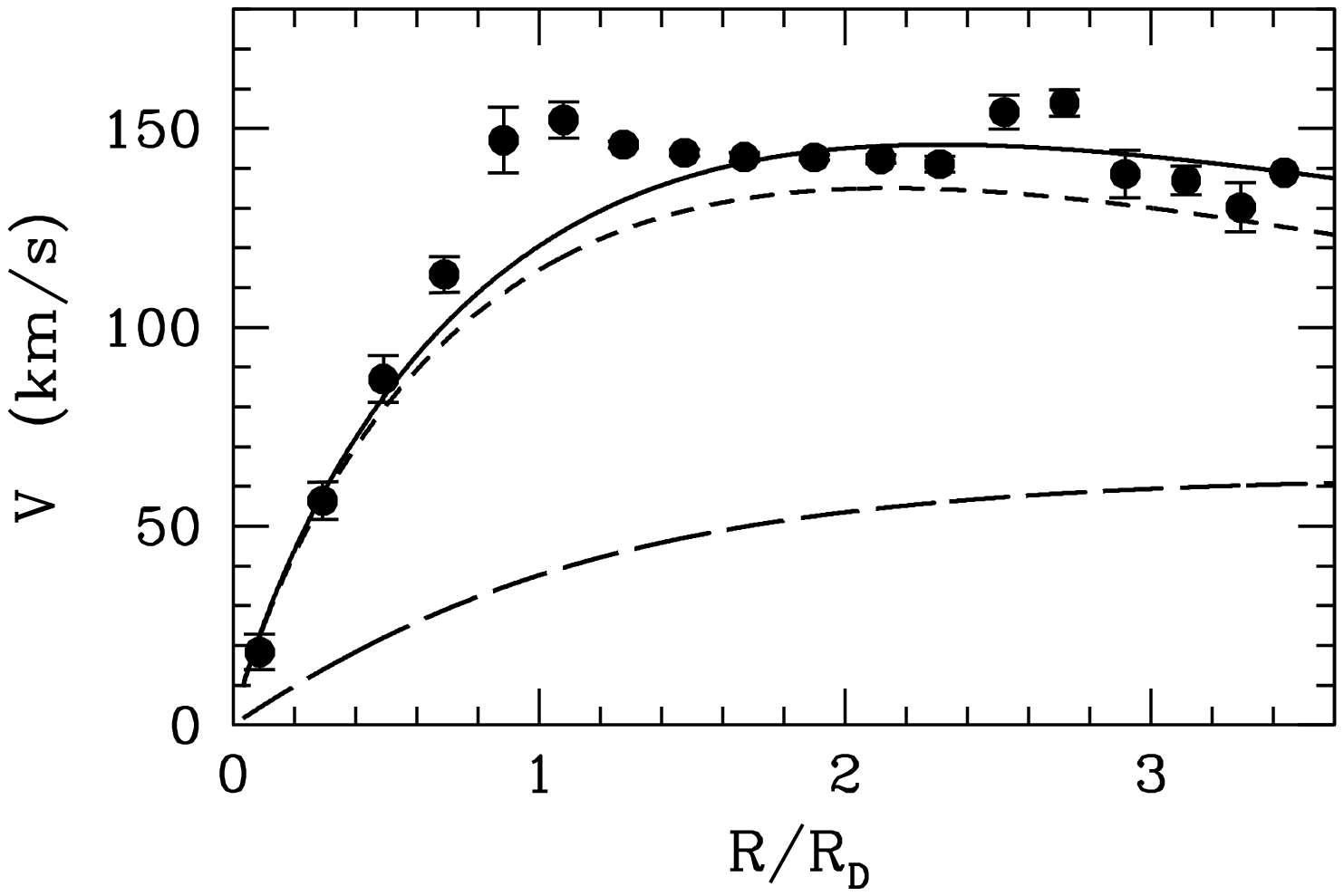}

UGC 6351\tabularnewline
\end{tabular}

\caption{Mass models of the galaxies in  our sample.  Filled circles with errorbars - the RCs,
  short dashed line - the contribution of the stellar disc, long
  dashed line - the contribution of the dark halo, solid line - the
  model circular velocity}
\end{figure}

Given the aim of this paper, it is worth representing  the dark matter
component with the simplest halo velocity profile $V_{h}^2 (r) $
(linked to the mass profile by $V_{h}^2 (r) = {G M_{h}(<r)\over
  r}~$):
$$
V_{h}^2(x) = V_{h}^2(1)  (1+ a^2)  x^2/(a^2 +x^2)
\eqno(2b)
$$
with $V_{h}^2(1) \equiv V_h^2(R_D)$ and $a$ free parameters.  The
above velocity profile implies a density profile with an inner flat
velocity core of size $\sim a R_D$, a constant central density and an
outer $r^{-2}$ decline, which, however is never reached in our RCs since 
generally  they do not extend  beyond $R_{last}\sim 3 R_D$. Let us  highlight that 
in the region in which most of the baryons lie and where we will measure 
the disk mass, eq(2b) can approximate, with proper 
values of the free parameters, a number of different halo distributions, 
including the NFW, the Burkert and the pseudo-isothermal ones. Obviously, 
for $R>R_{last}$, $V_{h}$ needs not to be represented by eq(2b).

The kinematical estimate of the disk mass $M_{kin} $
is obtained by fitting the observed rotation velocities  $V$ to the
model velocity curve  $V_{mod}$:
$$
V^2(x) = V^2_{mod} \equiv V^2_{d}(x,M_{kin}) + V^2_{h}(x; V_h(1), a)
\eqno(3)
$$
the model parameters, including the disk mass $M_{kin}$, are obtained by
minimizing the usual quantity (data-model)$^2$.

 An advantage  of this  method  emerges by noticing  that $M_{kin}$ are found very similar    
to the disk masses  computed by means of the
approach discussed above: a) by means of the equation: $|d\log ~ V_d(x)/d\log~ x
\simeq d\log V(x)/d\log~ x|< 0.05$ we   determine the inner
baryon dominance region (Salucci and Persic 1999), i.e.  the region
inside which the slope of the disk contribution to the circular
velocity coincides (within the observational errors) with the slope of
the latter; b) we   fit the RC's of this region with only the disk
contribution.   Since in most cases this region extend out to $R_D$,  we can write
  $M_{kin}\simeq G^{-1} V^2(1) (I_0K_0-I_1K_1)|_{0.5} R_D $: it is obvious that 
  the "theoretical"   uncertainties on  the kinematical estimates of the disk masses   are modest
  when the RCs show an inner region which is  well reproduced by the stellar component alone.   
   
The RC fits are excellent,  as those obtained  in Ratnam \& Salucci (2000) 
and Salucci et al (2000), showing that within the inner parts of spirals,  light
traces the dynamical mass. Thus,  the inner RC of spirals are reproduced by
just a stellar disk with a suitable choice of  its mass-to-light ratio. 
 In this way we  get the value $M_{kin}$ for the disk mass and the
formal uncertainty $\sigma_{kin}$ on $\log \ M_{kin} $,  found to
range between 0.10 and 0.20 dex,  and  mostly due to the (small but non zero)  observational error in the RCs 
and in the estimate of $R_D$. The resulting values for the disk masses are given in Table 1.

\begin{figure}
\vskip -0.4cm
\centerline{
\psfig{file=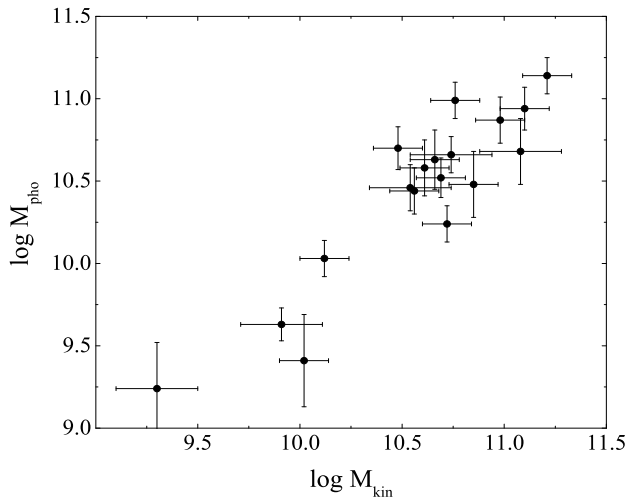,width=9.4truecm,height=7.8truecm}  
}
\vskip -0.7cm
 \caption{The $log \ M_{kin}$ vs $ log \ M_{pho}$ relationship.}
 \vskip -0.2cm 
\label{md2}
\end{figure}

A  further support for the reliability of  the above estimates   emerges by the fact that
 the   values  of $ log \ M_{kin}$   are found   to be within 0.1 dex  of 
those  computed by means of the following 
approach: a) with  the equation: $|d\log ~ V_d(x)/d\log~ x
\simeq d\log V(x)/d\log~ x|< 0.05$ we   determine the inner
baryon dominance region (Salucci and Persic 1999), i.e.  the region
inside which the slope of the disk contribution to the circular
velocity coincides (within the observational errors) with the slope of
the latter ; b) we   fit the RC's of this region with only the disk
contribution.  Then, it is very likely  that  the  {\it kin}  method is not significantly affected by
 model dependent    uncertainties.

\section{The Spectro-Photometric method}

We infer stellar masses from multi-color photometry following Drory,
Bender, \& Hopp (2004).  We compare multi-color photometry to a grid
of stellar population synthesis models covering a wide range in star
formation histories (SFHs), ages, burst fractions, and dust extinctions. 
We use photometric data in the ugriz bands from the Sloan Digital Sky 
Survey (SDSS) Data Release 4, augmented with  JHK data from the 2 Micron
All Sky Survey. We perform matched-aperture photometry with apertures
defined in the SDSS r-band image to obtain integrated galaxy colors to
the Petrosian radius. Photometric errors are 0.03-0.08 in J, H, and K,
respectively, $\sim$~0.1~mag in the u band, and $\sim$~0.01~mag in the
g, r, i, and z bands. Just as  a reference of the galaxy luminosities,
we  give  in Table 2 their absolute B-band magnitudes. 

Our stellar population model grid is based on the Bruzual \& Charlot 
(2003) stellar population synthesis package. We also use an updated 
version of these models (G. Bruzual, private communication). 
We parameterize the possible SFHs by a two-component model: 
a main component with a smooth analytically described SFH, and, 
superimposed, a short recent burst of star formation. The main 
component has a star formation rate of the form $\psi(t)
\propto \exp(-t/\tau)$, with $\tau \in [0.1, \infty]$~Gyr and a
metallicity of $-0.6 < \mathrm{[Fe/H]} < 0.3$.  The age, $t$, is
allowed to vary between 0.5~Gyr and the age of the Universe
(14~Gyr). We superimpose a burst of star formation, modelled as a
constant star formation rate episode of solar metallicity and of
100~Myr duration. We restrict the burst fraction, $\beta$, to the
range $0 < \beta < 0.15$ in mass (higher values of $\beta$ are
degenerate and unnecessary since this case is covered by models with 
a young main component). We adopt a Chabrier (2003) initial mass
function for both components. The main component and the burst are
allowed to independently exhibit a variable amount of extinction by
dust.  This takes into account the fact that young stars are found in
dusty environments and that the starlight from the galaxy as a whole
may be reddened by a (geometry dependent) different amount.

We compute the full likelihood distribution on a grid in this
6-dimensional parameter space ($\tau, \mathrm{[Fe/H]}, t, A_V^1,
\beta, A_V^2$), the likelihood of each model being $\propto
\exp(-\chi^2/2)$. In each object, we compute the likelihood
distribution of the stellar mass-to-light ratios that will give the
best estimated value $M_{pho}/L$, by weighting the mass-to-light
ratio relative to a spectro-photometric model and marginalizing  it over 
all stellar population parameters. The uncertainty in the derived 
$M_{pho}/L$ , and hence in stellar mass, is obtained from the width 
of this distribution and given in Table 2. 
While estimates of stellar population parameters 
such as the mean age, the star formation history, the burst fraction, 
and the dust content are subject to degeneracies and often are poorly 
constrained by the models, the value of the stellar disk mass obtained 
by marginalizing over the stellar population parameters is a lot more robust. 
On average $ \sigma_{pho} $, the width of the distribution of the likelihood 
of $M_{pho}/L$, at $68 \%$ confidence level is between $0.1$ and $0.2$~dex.
 The uncertainty in the estimated  stellar mass is mostly "theoretical";
it has a weak dependence on the
stellar mass itself (in that it increases with lower $S/N$ photometry)
and much of the variation of the errors is in spectral type:
early-type galaxies have more tightly constrained masses than late
types because their star formation histories are tightly constrained
while the ones of late-type galaxies are less well constrained due to
degeneracies with age and recent burst fractions. This dependence of
the uncertainties on spectral type is the dominant source of
uncertainty in the photometric mass of our sample. The contribution to
the uncertainty due to photometric errors is negligible in our
relatively high $S/N$ photometry, however,  about 20\% of the uncertaity
is due to errors in the determination of the extrapolated total
magnitudes and colors.

Note that masses computed with the BC07 models are lower by 0.1 to
0.15~dex compared to the ones using the BC03 models. This is due to
the higher red and infrared luminosities of intermediate age
($\sim$~0.8-2~Gyr) stellar populations in the newer models which owing
to a larger contribution of post-AGB stars in the newer models. This
particularly affects our sample of mostly relatively late type spiral
galaxies with extended star formation histories and significant recent
star formation.

\section{Results}

The two different estimates of the disk masses are shown in Figure 3.
A correlation yields:

$$
\log ~ M_{pho}=  (-0.4 \pm 1.27) +  (1.02 \pm  0.12) \log ~
M_{kin} 
\eqno(4)
$$
with a r.m.s of 0.23 dex.  By considering the errors on the separate
determination of $M_{kin}$ and on $M_{pho}$ the slope and zero-point
of the relation are consistent respectively with 1 and 0.  From Fig.
(3) and eq. (4) it is evident that the two estimates are {\it statistically}
equivalent, i.e.  {\it on average}:   $M_{pho}= M_{kin} $. Within a small scatter  
both mass estimates are suitable measures of the true disk (stellar)
mass. On an  individual basis, however,  the match can be  less impressive  given  the  presence of outliers up  to 0.5 dex off
the relationship.

 Eq(4),  showing that   the value of the   slope of relationship    is near to unity,  implies that the slope  of any relation between a  quantity $Q$  and the disk mass  $M_{true}$,  given  by  $log \   Q= const+ \alpha_{true}  \  log M_{true}$ does not change when we   substitute   the unknown $M_{true}$ values  with the available  $M_{pho}$ ones.
Moreover, considering that the  estimated  uncertainty on   $log ~ M_{pho}$ is about 0.25 dex  (a quantity much lesser than its variation among spirals), we can  substitute $M_{true}$   in the same way,  in order to derive  quantities  such as   the disc  mass function,  the halo-to-disc mass ratio or in statistical investigations of the Tully Fisher.   
\begin{figure}
\begin{center}
  \psfig{file=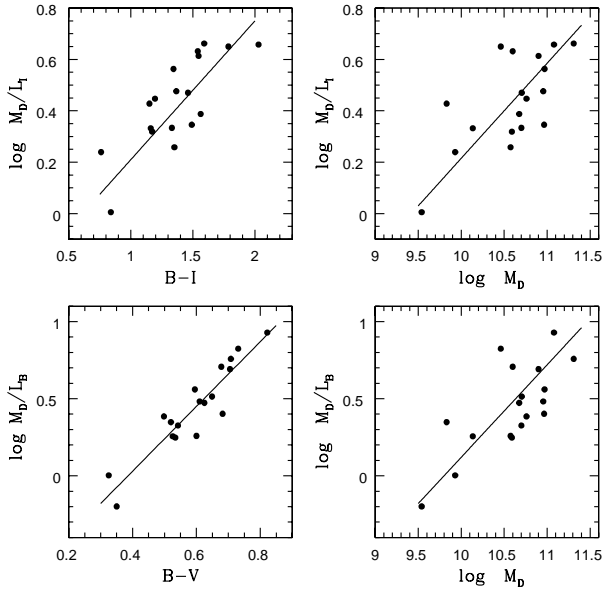,width=8.4truecm}  
\caption{Mass-to-light ratios vs color and disk mass} 
 \end{center}
 \vskip -0.3cm
\end{figure}

\section{The mass-to-light ratios in spirals}

Biases, observational errors and systematics of the two determinations
are independent, therefore we can define as an accurate measure of the
disk mass $M_D$  as the log average of the two different estimates:
$$
\log M_D \simeq \frac{1}{2} (\log M_{pho} +  \log M_{kin})
\eqno(5)
$$

From this we compute the spiral mass-to-light ratio. (AB magnitudes in the Johnson filters).
We find that this quantity, in spirals,  ranges over 1.2  dex in the B band and 0.7 dex in the I band
and, not unexpectedly,  depends  on the galaxy  broad-band  color.  In fact, we find the relation (see Fig. 4)

$$
 M_D/L_B= (0.66 \pm 0.13)   \times 10^{(2.1 \pm 0.3) (B-V) -0.3}
 \eqno(6a)
$$
that,   within the very small r.m.s of about 0.05 dex,   
reflects the fact that   older stellar populations are redder and have higher mass-to-light ratios.  
Similarly,  we get 

$$
 M_D/L_I= (1.6 \pm 0.3)   \times 10^{(0.54 \pm 0.1) (B-I) - 1}
 \eqno(6b)
$$
(about r.m.s. = 0.07 dex)

The agreement between the two determinations  allows us  to establish,  from the I
luminosity and B-I  broad band colors,   a solid {\it statistical}  estimate of the stellar mass
of spirals (i.e. well  within an uncertainty of 0.1 dex), that  may be  reduces  further reduced
 when the whole SED is considered.

Although not derivable from basic stellar physics as the previous ones,  we find that the mass to light 
ratio correlates with stellar mass (or luminosity) see Fig 4,
$$
 M_D/L_B=  (0.6 \pm 0.18) \times (M_D/10^{9} M_\odot)^{0.6 \pm 0.2}  
 \eqno(7a)
$$
within a r.m.s.  of 0.12  dex   and 

$$
 M_D/L_I=  (0.7 \pm 0.25) \times (M_D/10^{9} M_\odot)^{0.35 \pm 0.1}
 \eqno(7b)
$$
within a r.m.s. of about 0.15 dex.

\section{Discussion}

In this paper we find that the two main methods to measure the disk
masses in spirals, namely the SED and the RC fitting are both robust,
solid and consistent. In an illustrative way, in bulge-less systems, already 
the stellar disk  mass estimates  $0.66 \   L_B  \times 10^{(2.1  (B-V) -0.3)} $and $f \  G^{-1} 3.6 \  V^2(R_D)   R_D $ 
that we obtain from a  simplified implementations of the two methods are  equivalent and  reliable.

On the other hand, the agreement between the two methods implies a  support for a)
the existence of a Inner Baryon Dominated region, inside which the
stellar disk saturates the gravitational potential overwhelming that of  the DM
halo b) the assumed IMF and SFR histories: significantly different
choices would lead to $M_{pho} \neq M_{kin}$.

Our results confirm in a substantial way the work  by  Bell and de Jong (2001): they found within reasonable assumptions,  that the  stellar population models predict a strong correlation between  a colour of a stellar population and its stellar M/L ratio;   moreover, it emerged that the  slope of such relationship  is quite insensitive to the Star Formation Histories  and  the  mean ages of galaxies. For  the present sample we   find  (see figure 3)   statistically relevant  mass-to-light vs color relationships, with  values of their  slopes  ($~$2.1 and  $~$0.54 in the B and I band respectively)   in good agreement with  those  of  Bell and de Jong (2001) ($~$1.8 and  $~$0.6). In addition,  from the  compatibility of the  stellar population models with  the Tully-Fisher relation,  $log L =a + b \ log V$,  Bell and de Jong claimed  "maximun disk" mass distributions.  In spite of the fact that the TF relation is  strongly  biased by the  DM in a luminosity and radial dependent way (Yegorova et al, 2007),  our results  support such view: if  {\it on average} we assume the  "photometric" values for the  disk masses and  we  mass model the RC's,  we obtain a dark-luminous mass  decomposition  very similar to the "maximum disk" one.

The derived   values of the disk masses  imply  that spiral galaxies, unlike ellipticals, have a quite wide range in the
mass-to-light ratios, i.e.  almost a dex, reflecting an intrinsic spread in the  ages of their average stellar populations, confirmed  by their spread in colors. Moreover, it is evident that spiral disks are significantly less
massive than the elliptical spheroids of the same luminosity.   
  For ellipticals we have: $M_{sph}/L_B \sim 4\
\lambda^{0.2}$, $\lambda =L_B/ (2 \ 10^{10} L_{B\odot})$ (e.g. Borriello
et al 2000), with $0.5\leq \lambda \leq 10$, while for spirals we
have found (in section 3): $M_{D}/L_B \sim 2 \ \lambda^{0.6}$ with
$0.01 \leq \lambda \leq 5 $. In the luminosity range where ellipticals
and spiral coexist, spheroids are therefore more massive by an amount
$1.5 -2.5$ than disks of  the same luminosity.

 Finally,  there is  one case in which   $M_{pho}$ is too uncertain to  substitute  the   disk mass $M_{true}$ value: the DM  halo cusp-core controversy. In fact,  if in the    kinematical  mass modelling of  a   RC  we assume    $M_{true} =M_{pho}$,  we also  introduce in this  crucial constraint an  error of a size ranging   from  $ -0.23$ dex  to $  +0.23$ dex;  this  triggers a  serious uncertainty  and a  troublesome not-uniqueness in the RC's best fit solutions. One practical  example  is  ESO 287-G13 in Gentile et al 2004, according to the results of the present work  
the  {\it pho} estimate of the I-band  mass-to-light ratio of this object  ranges   between 0.6 and 2.3.   However, while the lowest  value allows    a stellar disk plus a   gaseous disk and a  NFW halo,  the highest value, instead,    leads  the above  model   to be  strongly  inconsistent  with the RC.   

After the results of this  pilot study  we can claim that,  by measuring the  disk  mass for  a  reasonable  number of objects (N $~ 100$) by means of  {\it both}  high quality  kinematics and extended SEDs,  it will be possible  1) to  establish an accurate   color vs disk mass  relationships that, when applied to very large samples ($N >1000$)  will give a reliable (for a number of issues) measure of the latter with    much  less observational effort than any  other method  2) to use the agreement of the two different measures as an observational tool to investigate the SFH and  the IMF of spirals. 

\begin{table}
\hspace{-0.4cm}
\begin{tabular}{lllllll lr}
Name  &$V_d$ &$M_{kin}$ &$dM_{kin}$&$M_{pho}$&$dM_{pho}$&$M_{B}$&ref& \\
             &     &   \\
UGC3944&100&  10.12& 0.12& 10.03& 0.11&-19.82& 1&\\
UGC4580&135&  10.48& 0.12& 10.7&  0.13&-21.31& 1&\\
UGC7549&62&    9.3 & 0.2&   9.24& 0.28&-19.72& 1&\\
UGC10706&208& 11.08& 0.2&  10.68& 0.2&-21.26&  1&\\
UGC10815&229& 11.1& 0.12&  10.94& 0.13&-20.99& 1&\\
UGC12354&100& 10.02& 0.12&  9.41& 0.28&-20.13& 1&\\
UGC5631&90&    9.91& 0.2&   9.63& 0.1&-19.53&  1&\\
UCG12810&220& 11.21& 0.12& 11.14& 0.11&-21.84& 1&\\
UGC5715&190&  10.76& 0.12& 10.99& 0.11&-21.8& 1&\\
UGC8460&145&  10.56& 0.12& 10.44& 0.14&-20.65& 2&\\
UGC4275&190&  10.85& 0.12& 10.48& 0.2&-21.15&  2&\\
UGC7823&165&  10.66& 0.12& 10.63& 0.18&-20.86& 1&\\
UGC8749&170&  10.72& 0.12& 10.24& 0.11&-20.32& 1&\\
UGC9598&160&  10.69& 0.12& 10.52& 0.12&-20.93& 1&\\
UGC9745&155&  10.61& 0.12& 10.58& 0.17&-21.01& 1&\\
UGC10545&155& 10.98& 0.12& 10.87& 0.14&-21.46& 1&\\
UGC4119&290&  10.74& 0.2& 10.66&  0.11&-20.43& 3&\\
UGC6351&220&  10.54& 0.2& 10.46&  0.14&-19.19& 3&\\
\end{tabular}
\caption{Columns:  1- name of the galaxy,
	2- reference disk velocity $\equiv {1\over 2}  G M_{kin}/R_D$ , 
3- kinematical  mass (log),  4 - its uncertainties, 5- spectro photometric 
mass (log), 6 -  its uncertainties, 7 - absolute B-band magnitude 
(HyperLeda database), 8- RC references: Courteau-1, Vogt-2, our data   
(Asiago Observatory)-3}
\vspace{0.5cm}

\end{table}

\section {Acknowledgments}

We thank Alessandro Pizzella for  helpful discussions  and for  the data of Sa galaxies, Ana Babic and Michael Cook  for helpful discussions and the anonymous referee for comments that have improved the presentation of our results.

\appendix


\label{lastpage}

\end{document}